\pdfoutput=1

\documentclass[11pt]{article}

\usepackage{EMNLP2023}

\usepackage{times}
\usepackage{latexsym}

\usepackage[T1]{fontenc}

\usepackage[utf8]{inputenc}

\usepackage{microtype}

\usepackage{inconsolata}

\usepackage{graphicx}

\usepackage{listings}
\usepackage{tikz}
\usepackage{pgfplots}
\usepackage{comment}
\pgfplotsset{compat=1.18}

\usepackage[title]{appendix}

\newcommand\siwarex{\textit{siwarex}}

\usepackage{verbatimbox}

\title{Declarative Techniques for NL Queries over Heterogeneous Data}

\author{Elham Khabiri, Jeffrey O. Kephart, Fenno F. Heath III, Srideepika Jayaraman \\ {\bf Fateh A. Tipu}, {\bf Yingjie Li}, {\bf Dhruv Shah}, {\bf Achille Fokoue}, {\bf Anu Bhamidipaty} \\ IBM Research, Yorktown Heights, NY 10598 USA 
}

\begin{document}
\maketitle

\begin{abstract}

In many industrial settings, users wish to ask questions in natural language, the answers to which require assembling information from diverse structured data sources. With the advent of Large Language Models (LLMs), applications can now translate natural language questions into a set of API calls or database calls, execute them, and combine the results into an appropriate natural language response.
However, these applications remain impractical in realistic industrial settings because they do not cope with the data source heterogeneity that typifies such environments. In this work, we simulate the heterogeneity of real industry settings by introducing two extensions of the popular Spider benchmark dataset that require a combination of database and API calls. Then, we introduce a declarative approach to handling such data heterogeneity
and demonstrate that it copes with data source heterogeneity significantly better than state-of-the-art LLM-based agentic or imperative code generation systems. Our augmented benchmarks are available to the research community.

\end{abstract}

\section{Introduction} 

In many industrial settings, users wish to ask questions whose answer may be derived from structured data sources such as a spreadsheets, databases, APIs, or a combination thereof. Often, users don't know whether the answer exists anywhere in the system, or if so how to identify or access the right data source. In recent years, a variety of applications and research efforts have been developed to address this issue.  As a result, multiple benchmarks~\cite{yu-etal-2018-spider,zhong2017seq2sql, li2023can} and systems ~\cite{gao2023texttosqlempoweredlargelanguage, deng2025reforcetexttosqlagentselfrefinement} have been developed to improve the capability of LLMs to successfully convert natural language questions into SQL queries against a database (the text-to-SQL problem). Many other benchmarks~\cite{patil2023gorilla, li2023api} and approaches ~\cite{jha2025itbenchevaluatingaiagents, prabhakar2025apigen} have targeted the ability of LLMs to sequence and invoke the right APIs to answer a user’s question.

Such benchmarks have driven steady progress towards ready and intuitive access to proprietary sources of information via either SQL or API calls, but they fail to test the ability of systems to cope with a combination of the two. In practical applications, one frequently encounters questions like ``Which Xylem pumps at Bedford have experienced anomalous temperatures today?'', which requires combining a database call to retrieve pumps with the right manufacturer and location with an API call to sense or compute temperature anomalies.\footnote{Additional examples of such questions are provided in Appendix~\ref{sec:IndustrialExamplesAppendix}.} 


In this paper, we explore the problem of data heterogeneity.  While existing agent-based architectures (e.g., ReAct~\cite{yao2022react}) can dynamically orchestrate retrieval and aggregation of information across APIs and databases, they tend to be brittle, expensive to run, and difficult to scale in production. A core limitation of such approaches is that they conflate representation of the user's intent with planning an efficient execution sequence into a single step that is handled directly by LLMs.

We present a more practical
architecture that retrieves and aggregates data from databases and APIs by cleanly separating a representation of the user's intent from planning an efficient execution sequence. We rely on SQL as a declarative language that expresses the user's intent and use User Defined Functions (UDFs) to invoke APIs from within SQL queries. By leveraging the UDF capability of modern database systems (e.g., postgres or DB2) to invoke external APIs, we place APIs on the same footing as database tables. In so doing, we leverage decades of research in SQL query optimization and planning for efficient orchestration and aggregation across both database tables and APIs (through their corresponding UDFs). We also explore an imperative approach that uses an LLM to generate imperative python code that stitches together information from heterogeneous sources and then executes the generated code. 

Since we are unaware of any existing benchmark against which we can compare our declarative approach against imperative or agent-based approaches, we have created two new benchmarks consisting of questions whose answers require a combination of database and API calls, both of which are augmentations of the popular Spider dataset and benchmark. Benchmark I replaces a fraction of the real Spider database tables with equivalents that are executed via APIs. This allows us to directly test the mechanism by which database and API calls are combined without having to change the questions or their ground-truth answers from the original Spider benchmark. Benchmark II introduces a new set of scalar APIs that perform simple lexical, numeric, or geo-spatial operations. From a subset of two dozen Spider databases, we transform questions from the original Spider database into new questions that require interleaving database operations with compositions of 1-3 scalar APIs. We establish a set of corresponding ground-truth answers through a semi-automated process that generates over 2300 human-vetted question/answer pairs.
 
In the remaining sections, we briefly survey related work, detail our implementation of the systems that we are comparing, introduce the benchmark datasets, and summarize experiments that establish that our proposed approach outperforms imperative and agent-based approaches in several key metrics. Our main contributions include:

\begin{enumerate}
    \item Two new benchmarks for assessing the ability of LLM-based systems to cope with data source heterogeneity;\footnote{These are available at \url{https://huggingface.co/datasets/ibm-research/SQL-API-Bench}.}
    \item A declarative approach that uses SQL statements to represent user intent and leverages User-Defined Functions (UDFs) to place external APIs on the same footing as database tables, allowing them to be manipulated by standard database execution engines; and
    \item Empirical evidence that our approach significantly outperforms an imperative code generation approach and an agent-based approach that combines state-of-the-art Text-to-SQL and API calling tools.
\end{enumerate}

\section{Related work} 
\label{sec:relatedwork}


We segregate previous work into three categories: structured data retrieval, LLM-based tool-calling, and coping with data heterogeneity. While there is much good work on systems that cope with Text-to-SQL and tool/API-calling individually, and some initial efforts to address data heterogeneity, there remain significant gaps in techniques that bridge across heterogeneous data sources and benchmarks that measure their efficacy.

\subsection {Structured data retrieval}
Popular large-scale Text-to-SQL benchmark datasets consisting of thousands of pairs of questions and their corresponding SQL ground truth include Spider \cite{yu-etal-2018-spider}, Spider-2 \cite{lei2025spider20evaluatinglanguage}, and BIRD \cite{li2024can}. Enterprise deployment of Text-to-SQL systems faces significant challenges, as
they must handle massive schemas containing over 1,000 columns, support multiple SQL dialects, and accommodate complex analytical requirements including data transformations and advanced analytics. Recent work like ReFoRCE ~\cite{deng2025reforcetexttosqlagentselfrefinement} and  DAIL-SQL~\cite{gao2023texttosqlempoweredlargelanguage} have addressed many of these issues and achieved top performance on the Spider Text-to-SQL benchmark. However, these techniques apply strictly to structured database queries, and cannot handle user requests requiring external API calls.


\subsection{LLM-based tool calling}
The Berkeley Function-Calling Leaderboard (BFCL)~\cite{patil2025bfcl} maintains a leaderboard for state-of-the-art systems in the tool-calling task. Toolformer~\cite{schick2023toolformer} and ToolLLM~\cite{kojima2023toolllm} use LLMs to select and invoke the right APIs from an available set. \cite {elder2025invocableapisderivednl2sql} use the {\em structure} of SQL to select and orchestrate APIs according to their primary role, e.g. selection or filtering, but the end result is only a set of structured API calls that do not access databases. 
NESTFUL~\cite{basu2025nestfulbenchmarkevaluatingllms} presents a benchmark for evaluating LLMs on nested sequences of API Calls, which provides a harder task than calling individual APIs, but it sticks to the structured nature and singular modularity of APIs.
XLAM~\cite{zhang2024xlamfamilylargeaction} contains a set of fine-tuned LLMs that choose and execute API calls. Currently leading the BCFL leaderboard, it is regarded as the state-of-the-art in multi-turn tool calling. However, as it is specifically trained for a structured, function-calling environment, its support for multi-modal reasoning is limited. Moreover, changes to the set of tools or their interfaces requires re-training or fine-tuning~\cite{lin2024hammerrobustfunctioncallingondevice}.

\subsection{Handling heterogeneous data sources}

NL2Code has emerged as an efficient way to handle complex workflows, including data retrieval from multiple sources. CodeAct~\cite{wang2024executablecodeactionselicit} generates executable Python code that serves as a unified action space for combining tool use, control flow, and data handling. However, typically, such systems fail in complex, heterogeneous scenarios that require multi-step reasoning workflow to combine the data.


BLENDSQL~\cite{glenn-etal-2024-blendsql} is a hybrid dataset that introduces new SQL extensions (LLM functions) that support hybrid question and answering across structured data (tables) and unstructured data. In contrast, we rely on the strength of existing SQL dialects that support user-defined functions. We believe that our approach could achieve the same effect as BLENDSQL without requiring special extensions, simply by positioning the unstructured data retrieval APIs as UDFs.

\section{Approaches to handling heterogeneity} 
\label{sec:approaches}
\label{sec:framework}





Here we describe three approaches for handling NL queries over heterogeneous data sources. The Declarative and Imperative NL query strategies use LLM-based code generation, while the Agentic query strategy is a ReAct agent-based solution that relies on specialized tools to access APIs and databases. For consistency, and due to its proven effectiveness as a component of Text-to-SQL systems ~\cite{gao2023texttosqlempoweredlargelanguage, deng2025reforcetexttosqlagentselfrefinement}, all three query strategies employ the Mistral-Large LLM~\cite{Mistral-Large}.

\subsection{Declarative approach}
\label{sec:DeclarativeApproach}

\begin{figure*}[hbtp]
\centering
   \includegraphics[scale=0.50]{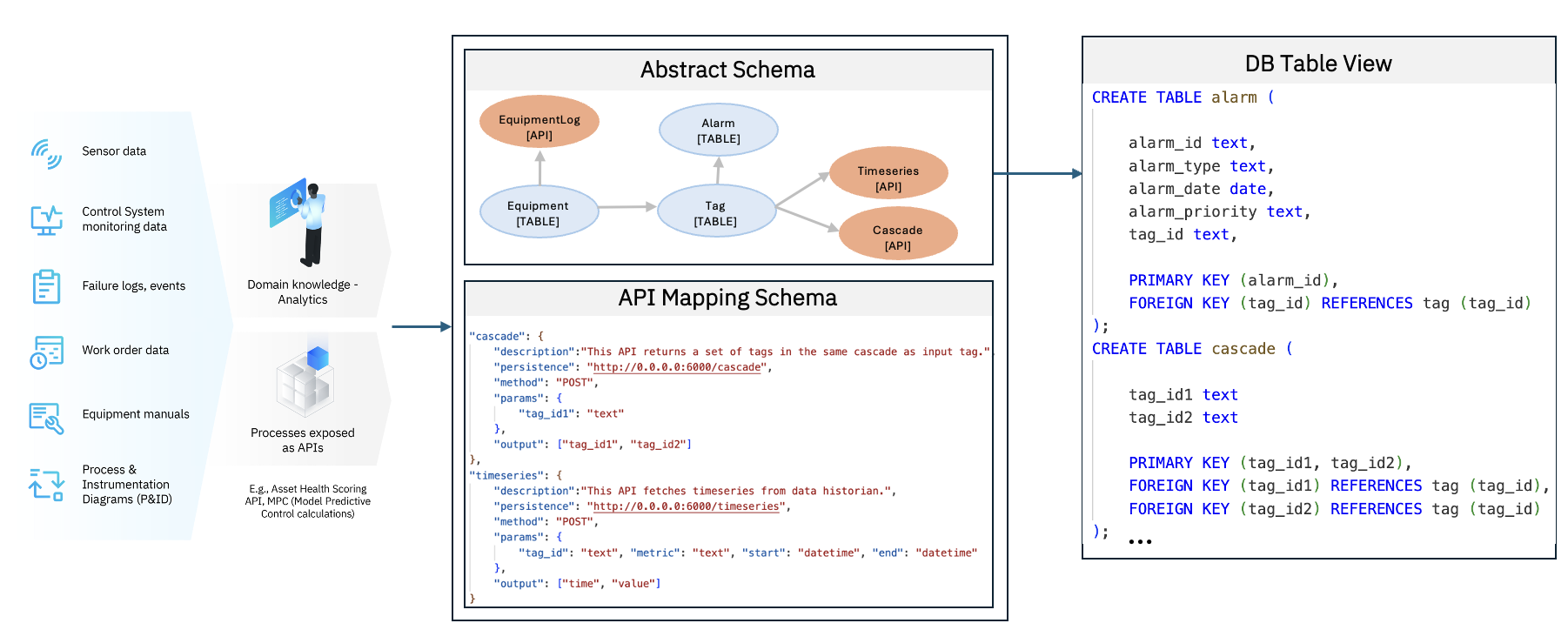}
    \caption{\textbf {Example of schemas and table view used by \siwarex.} The Abstract Schema and API Mapping Schema required by {\em \siwarex} can be provided manually or extracted from domain metadata. If a database schema is provided, the Abstract Schema can be extracted from it automatically; likewise the API Mapping Schema can be extracted automatically from an OpenAPI spec. For systems that mix DB access and API calls, the edges between API and DB nodes in the Abstract Schema may be augmented by a minimal amount of expert knowledge. Once the Abstract Schema is created, a relational schema (DB Table View) is generated from it automatically. The DB Table View represents all entities consistently as tables regardless of whether they are actually tables or APIs.}
    \label{fig:SchemaExtraction}
\end{figure*}

This section describes the components of our declarative \textit{\siwarex} framework, which leverages two data source schema that can be provided by users or derived semi-automatically from domain metadata:
\begin{enumerate}
    \item The \textit{Abstract Schema}, in the form of an Entity-Relationship Diagram, provides a global view of the data source properties and interrelationships in a format that is agnostic to whether the data source is a database table or an API. 
    \item The \textit{API Mapping Schema} provides information necessary to invoke an API call, such as the URL, the method (POST, GET, etc.), and details of the input and output parameters.
\end{enumerate}
As illustrated in Figure~\ref{fig:SchemaExtraction}, the Abstract Schema and API Mapping Schema are generated either manually or via an automated process that leverages database schemas, OpenAPI specs, or other metadata. A deterministic offline process automatically converts the Abstract Schema into a relational schema that is used at runtime to generate SQL queries from NL questions. In that relational schema, tables corresponding to APIs (e.g., cascade) are \textit{virtual tables}, each of which is associated with a corresponding User Defined Function (UDF) that invokes the associated API by consulting details provided in the API Mapping. An example UDF and its associated API wrapper are illustrated in Appendix~\ref{sec:UDFAppendix}.

\textit{\siwarex} includes two key components:

\begin{itemize}
    \item A standard \textbf{Text2SQL} module that, given the relational schema generated from the Abstract Graph (i.e., with virtual tables) and a user's NL question, generates a corresponding SQL. An example is provided in Appendix~\ref{sec:DeclarativeExample}.

        
    \item A rule-based \textbf{Query Rewriter} that bridges the physical and logical representation of data entities. It rewrites LLM-generated SQL containing \textit{virtual tables} into executable SQL by replacing \textit{virtual tables} with their corresponding UDFs, and (based on static SQL analysis) passes the right arguments.
\end{itemize}

We also implemented a variant of the Declarative method called \textbf{Declarative2} that can be used in situations where the UDFs represent scalar functions. Instead of using a query rewriter and treating UDFs as virtual tables, each scalar API is wrappered as a UDF and called directly in a manner analogous to built-in SQL functions like LENGTH. An example is provided in Appendix~\ref{sec:Declarative2Example}. The resultant SQL expression tends to be easier to understand than that produced by \textbf{Declarative}, but its applicability is more limited. It is only possible to evaluate it for Benchmark II (described in Section~\ref{sec:BenchmarkII}).


\subsection{Imperative approach}
\label{sec:ImperativeApproach}

We implemented an imperative approach that generates for each natural language question a Python program that choreographs the various database and API calls that are required to answer it. The relevant database schema are extracted dynamically from the database, and the relevant API specifications obtained. This information, along with the input question and sample rows from each table, is included in an LLM prompt that generates a Python program that is then executed to return the answer. An example of such an auto-generated Python program can be found in Appendix~\ref{sec:ImperativeExample}.

\subsection{Agentic approach}
\label{sec:AgenticApproach}


We implemented an agentic approach
that uses ReAct~\cite{yao2022react} reasoning to call various tools to answer a given NL question, including:
\begin{enumerate}
    \item SQLDatabaseToolkit, a Langchain toolkit that interacts with SQL databases, which we configure to use the Mistral-Large LLM~\cite{Mistral-Large}. 
    
    \item xLAM~\cite{prabhakar2025apigen}, an API-calling tool at the top of Berkeley API Leader board as of mid-2025.\footnote{gorilla.cs.berkeley.edu/leaderboard.html} From a set of API metadata provided to this tool, it selects those most likely to help answer the question.

\end{enumerate}

\noindent ReAct is widely regarded as a standard for evaluating tool-calling performance of LLMs~\cite{10.5555/3692070.3693047, Fu2024PreActPE, basu2025nestfulbenchmarkevaluatingllms}.

The agent is provided with relevant metadata that includes the names and descriptions of available APIs and table schema. On that basis, it generates the sequence of APIs and database queries that must be executed to answer the given question. An example chain-of-thought trace is provided in Appendix~\ref{sec:AgenticExample}.

    

\section{Two benchmark datasets} 
\label{sec:benchmarks}
\label{sec:dataset}
To assess the ability of any system to cope with heterogeneous data sources, appropriate benchmarks are needed. Since we are unaware of any that exist, we have created two that we are sharing with the research community. One approach would be to collect questions from real industrial Q\&A examples, but sharing such a dataset openly would face practical and political obstacles. Instead, we opted to modify Spider, a popular Text-to-SQL benchmark~\cite{yu-etal-2018-spider} that consists of a collection of databases and tables plus several thousand pairs of natural language queries and their associated ground-truth SQL translations.
Our benchmarks augment Spider in two distinct ways.

\subsection{Benchmark I}
\label{sec:BenchmarkI}

Benchmark I leaves all natural language questions as is, but replaces a fraction of the Spider database tables with equivalent API calls. Since many natural language queries in Spider require combining information from multiple tables, replacing some tables with API calls necessitates combining database and API calls to answer a NL question. We use a subset of the full Spider benchmark containing 948 questions and associated ground truth SQLs.

For example,
the Spider dataset includes the database \textit{museum\_visit}, which contains 3 tables: \textit{museum}, \textit{visitor} and \textit{visit}. The \textit{museum} table has the following SQL definition:
\begin{lstlisting}
CREATE TABLE museum (
    Museum_ID int PRIMARY KEY,
    Name text,
    Num_of_Staff int,
    Open_Year text
);
\end{lstlisting}

\noindent To generate the API equivalent of \textit{museum}, we programmatically convert its SQL definition to an API \textit{/museum} that is written in Python using the Flask framework. When executed, the API collects all of the table records from the database into a dataframe. Then, it applies filtering, selection and aggregation operations (written in Python) to that dataframe to produce the same rows and columns that the SQL execution would have produced. Finally, \textit{/museum} is wrappered as a a user-defined function (UDF).

We also programmatically convert the SQL definition of \textit{museum} to a Swagger definition for \textit{/museum}:

\begin{lstlisting}
/museum:
    post:
      description: The API 'museum' handles requests 
      regarding 'museum_id, name, num_of_staff,
        open_year', in the context of %'museum_visit'.
      requestBody:
        required: false
        content:
         application/json:
            schema:
              type: object
              properties:
                museum_id:
                  type: integer
                name:
                  type: string
                num_of_staff:
                  type: integer
                open_year:
                  type: string
      responses:
        '200':
          description: Data returned.
\end{lstlisting}

Suppose a user asks \textit{``What are the opening year and staff number of the museum named Plaza Museum?''}. In the original Spider, this is converted to the SQL statement \textit{"SELECT Num\_of\_Staff , Open\_Year FROM museum WHERE name = `Plaza Museum'"}, which is then executed on the database. However, if the database table \textit{museum} is replaced by the API \textit{/museum}, the system must call the \textit{/museum} API with the parameter \textit{name} =``Plaza Museum''. Now suppose that the user asks: \textit{"What are the id, name and membership level of visitors who have spent the most money in total in all museum tickets?"}. In the original Spider, this would entail joining the \textit{museum} and \textit{visit} tables. In the extended benchmark, the system must combine results of a database call to the \textit{visit} table with results of an API call to \textit{/museum}.



\subsection{Benchmark II}
\label{sec:BenchmarkII}

Benchmark II augments Spider by introducing a set of 16 scalar APIs that perform lexical, numeric, or geospatial operations that are generic enough to ensure that they integrate naturally with most of the existing Spider questions. Examples include counting the number of syllables in a string; determining whether an integer is a prime, a square, or a Fibonacci number; determining the latitude, longitude, country, or province of a place; or calculating the distance between two places. The numeric APIs accept float or integer inputs; floats are truncated to integers. All geospatial APIs are wrappers around Google Geospatial APIs.\footnote{\url{https://developers.google.com/ar/develop/geospatial}}

Given a question/SQL pair from the original Spider benchmark, we prompted an LLM with the original Spider question/SQL pair, a set of schemas for the Spider and scalar APIs, and a request to appropriately blend 1-3 APIs into the original Spider question. We applied this transformation 3-5 times to each question in 26 selected Spider databases, resulting in 5456 candidate augmented questions.

Unlike the original Spider benchmark, Benchmark II does not provide a ground-truth SQL expression for each question, as the APIs that we have introduced have no SQL equivalent.
Instead, the ground truth consists of rows that represent the correct answer to the question. To generate these ground-truth rows, we used three different LLM-based techniques. One technique treated the API calls as virtual tables that were joined with the original Spider tables, while the other two executed the APIs directly as User-Defined-Functions (UDFs).

Out of the 5456 candidate
augmented questions, 1649 produced no rows for any of the techniques. In many of these cases, the ground-truth
SQL was legitimate, but no rows were produced due to the extra restrictions imposed by the scalar APIs. Since there are many erroneous ways to generate no rows, such questions would undermine the utility of the benchmark. Therefore, we dropped them from further
consideration.\footnote{An alternative would have been to augment the Spider tables to include
rows that match the extra criteria, but we
felt that deviating from the well-established Spider content would make our dataset less useful to the research community.} Each of the remaining 3807 question/answer pairs was subjected to a human vetting process that consumed over 100 person-hours, resulting in the final set of 2338 question/answer pairs that comprise Benchmark II.

This benchmark creation procedure yielded questions that, while decidedly nerdy in nature, serve our purpose of requiring a question-answering system to perform a mix of database and API calls.\footnote{Moreover, piggybacking on the existing Spider benchmark enables us to calibrate our methods against it and generate a dataset an order of magnitude larger than would have been feasible otherwise.}
For example, the original Spider utterance

\begin{quote}
``What are the names and birth dates of people, ordered by their names in alphabetical order?'' 
    
\end{quote}

\noindent becomes QuestionID \emph{poker\_player.165}:

\begin{quote}
``What are the names and birth dates of people who live in countries where the name has a prime number of syllables, ordered by their names in alphabetical order?''    
\end{quote}

\noindent Answering this question correctly requires calling the \emph{get\_country\_of\_place} API on the nationality field of the people table in the \emph{poker\_player} database, and then applying the \emph{count\_syllables} and \emph{is\_prime} APIs successively to that result.

Further details about the process by which Benchmark II was created are provided in Appendix~\ref{sec:BenchmarkIIAppendix}, and more examples of transformed questions are provided in Appendix~\ref{sec:TransformedExamplesAppendix}.

\section{Experiments} 
\label{sec:experiments}
\label{sec:experiments}

We conducted two sets of experiments: one using Benchmark I and the other using Benchmark II. In each case, we compared the accuracy of the three methods introduced in Section~\ref{sec:framework} and conducted further investigations to gain qualitative insights into what characteristics of the methods most contributed to their overall efficacy.

In both experiments, accuracy was based on retrieved data. For Benchmark I, the benchmark rows were generated by executing the original gold standard SQL on the original Spider database. For Benchmark II, the benchmark rows were obtained by the human-vetting process overviewed in Section~\ref{sec:benchmarks} and detailed in Appendix~\ref{sec:BenchmarkIIAppendix}. Returned rows were matched against the benchmark rows using 
the comparison approach introduced by~\cite{zhong2020}, which requires that the correct rows be retrieved but forgives extra columns. The reported accuracy was the ratio of questions for which the comparison was deemed a match.


\subsection{Benchmark I experiments}
\label{sec:BenchmarkIExperiments}

\begin{figure}[hbtp]
\centering
   \includegraphics[scale=0.30, trim = 1.65in 0in 0.5in 0.5in, clip]{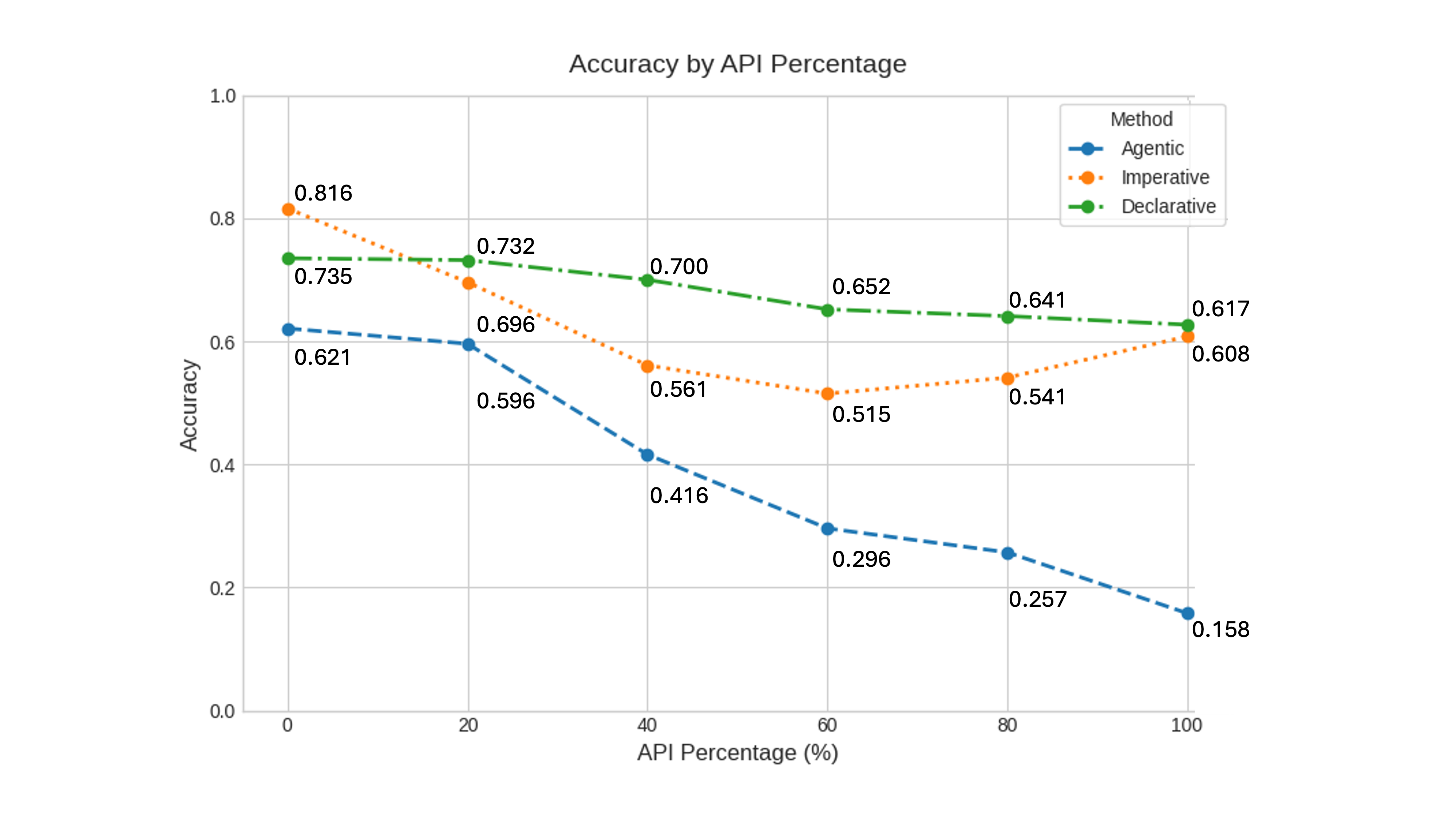}
    \caption{Measured accuracies for \textbf{Agentic} (xLAM-based), \textbf{Imperative} and \textbf{Declarative} methods vs the percentage of DB tables that were replaced by APIs.}
    \label{fig:Benchmark1_accuracy}
\end{figure}

Figure~\ref{fig:Benchmark1_accuracy} shows the accuracy of the methods introduced in Section~\ref{sec:framework} as a function of the percentage of database tables that have been replaced by APIs. 


Without any APIs (at 0\%), the Imperative approach outperforms the other approaches, achieving an accuracy of 0.816, which is comparable to state-of-the-art open-source LLMs on zero-shot evaluation.\footnote{See https://yale-lily.github.io/spider.} As the proportion of API calls increases, the accuracies of the Agentic and Declarative systems decrease monotonically, albeit much more steeply for Agentic than for Declarative. In contrast, the accuracy of the Imperative approach is {\em not} monotonic; it struggles most when the mixture of API and DB calls is roughly even. Error analysis reveals that, under these conditions, the complexity of the python code generated by Imperative is the greatest, and its accuracy is impaired by programming errors that arise when stitching together API calls and DB queries (typically inconsistent variable use).
For API percentages between 20\% and 80\%, the Declarative approach outshines the other two substantially. When all of the DB queries are replaced with API calls (100\%), Declarative is still the best, but only marginally better than Imperative. 


Careful error analysis reveals that three factors account for most of the Agentic method's poor performance for mixtures of DB and API calls:

\begin{enumerate}
    \item \textbf{Sequencing}. State-of-the-art tool-calling LLM based systems are not trained, fine-tuned or optimized to perform complex API sequencing, merging, and aggregation tasks. They perform relatively well on questions whose answers require a single API call, but at higher API percentages even the easier Spider queries (e.g., ``What is the total number of singers?'') typically require invoking multiple APIs and sequencing them properly (e.g. invoking getAllSingers followed by getSize).
    
    \item \textbf{Routing}. Even when the system generates a proper sequence, the master LLM sometimes fails to properly route the decomposed questions to the appropriate database or API tool. 
    
    \item \textbf{Hallucinated or improperly bound inputs}. Even when the right API is selected (e.g. {\em is\_prime}), the arguments for its invocation are often hallucinated or otherwise incorrect. For example, we have observed cases where the system correctly finds dozens of items with e.g. ``count > 3'' but then loses track of some of them, resulting in incorrect aggregation. 
\end{enumerate}

 The Declarative approach avoids all these issues by providing to an LLM a single relational view that removes all the complexity of dealing with multiple heterogeneous sources. To the LLM, everything appears to be relational, and thus it can leverage its Text-to-SQL capability to generate a SQL query for each NL question. The Query Rewriter is then responsible for injecting API invocations through UDFs with the proper arguments (inferred from a deterministic analysis of the SQL query).

\subsection{Benchmark II experiments}
\label{sec:BenchmarkIIExperiments}

Table~\ref{tab:Benchmark2_accuracy} summarizes the accuracy and the average execution time per question for the \textbf{Agentic} (xLAM), \textbf{Imperative}, \textbf{Declarative} and \textbf{Declarative2} approaches for the 2338 questions that constitute Benchmark II. 

Both declarative approaches are somewhat more accurate than \textbf{Imperative} and vastly more accurate than \textbf{Agentic}. Detailed comparisons of the 204 questions that were answered correctly by \textbf{Declarative2} but not by \textbf{Declarative} indicate that \textbf{Declarative} can suffer from errors that creep in during either the query rewriting or the extra virtual table joins. We believe that improvements to the query rewriter would close this gap somewhat. While \textbf{Declarative2} enjoys an accuracy advantage over \textbf{Declarative}, its use is limited to scalar APIs, whereas the \textbf{Declarative} method described in Section~\ref{sec:framework} applies broadly to APIs that generate outputs representable as scalars, vectors or tables.
The total execution time (including the time to process the NL into SQL or a program and execute the resulting database and API calls) for \textbf{Agentic} was notably slower than that of \textbf{Imperative} and the two declarative approaches. The trace shown in Appendix~\ref{sec:AgenticExample} suggests that loops over LLM invocations are one large source of inefficiency for \textbf{Agentic}.

\begin{table}
    \centering
    \begin{tabular}{|l|r|r|}
        \hline
         \bf{Method} & \bf{Accuracy} & \bf{Exec Time (sec)} \\
        \hline
         Agentic & 0.357 & 37.81 \\
         Imperative & 0.614  & 15.56 \\
         Declarative & 0.639 & 16.50 \\
         Declarative2 &  {\bf 0.689} & 16.24 \\
        \hline
    \end{tabular}
    \caption{{Accuracy and average execution time per question for the four query strategies on Benchmark II.}}
    \label{tab:Benchmark2_accuracy}
\end{table}

\section{Conclusion}

The ability to answer questions in industrial systems typified by data source heterogeneity is a critical and hitherto unmet need. To help address that gap, we introduced {\siwarex}, a declarative system that uses SQL to represent user intent in conjunction with virtual tables and UDFs, thereby enabling external APIs to be treated alongside database tables in a unified framework.
We also introduced two new benchmarks that assess a system's ability to cope with data heterogeneity and used them to establish the superiority of the declarative approach over imperative and agentic approaches. We have released these benchmarks to spur further research in this area.\footnote{Please visit \url{https://huggingface.co/datasets/ibm-research/SQL-API-Bench}.}



\section*{Limitations}
One limitation that we are eager to address in future work is that our benchmark's evaluation metric only considers the execution {\em accuracy} of the final results. Especially since we wish to produce a benchmark that meaningfully captures practical issues that arise in industrial settings, it is incumbent on us to augment this metric with an execution {\em performance} metric (i.e. efficiency or speed). While our augmentations of the Spider database had several advantages, in general the table sizes are too small (just dozens of rows) to support credible measurements of the system execution speed, both in terms of the time required to formulate the declarative statement and the time required to execute it. Augmenting a benchmark with larger tables, such as BIRD~\cite{li2023can}, would be far preferable for such a purpose.

Benchmark II only includes APIs that produce scalar outputs (e.g. Boolean outputs for APIs like \textit{/is\_prime}, or numeric outputs for APIs like \textit{/count\_syllables}). In many industrial use cases, APIs produce vector or table outputs. For example, the correlation between two sensors can be expressed as a numeric scalar, but correlations between one sensor and several others would naturally be expressed as a vector, and correlations among all sensor pairs would most naturally be represented as a table.  As mentioned in Section~\ref{sec:DeclarativeApproach}, the Declarative method is in principle capable of handling APIs that generate vector or table outputs. In future work, we hope to update Benchmark II to include such APIs and then measure the relative effectiveness of the Declarative, Agentic, Imperative approaches.

Finally, another important future extension is to deploy and test our declarative framework in an industrial system that contains heterogeneous data sources. This being the entire inspiration and motivation for our work, we are confident that our framework can be applied in systems that contain a variety of SQL and noSQL databases as well as APIs that access and/or analyze time series data. However, many practical questions remain to be answered, including the response time as experienced by an end user (which combines the formulation and execution times as mentioned above) and the degree to which creating API mappings for existing APIs can be automated to ensure that the system can be deployed quickly in new environments with minimal configuration.
 

\bibliography{main}
\bibliographystyle{acl_natbib}

\begin{appendices}

\section{Hybrid questions}
\label{sec:IndustrialExamplesAppendix}

Table~\ref{tab:industry_examples} of this appendix provides a few more examples of questions inspired by industry scenarios with which we are familiar, the answers to which require combining database and API calls.

\begin{table*}
    \centering
    \begin{tabular}{|p{2.5in}|p{2.5in}|}
        \hline \bf{Question} &
         \bf{Required DB and API calls} \\

        \hline
        What is the condition of all submersible pumps in my organization? & 
         \begin{itemize} 
         \item DB call to retrieve assets with type = `submersible pump' and owner = `me'
         \item API call to condition insight analyzer that either applies textual analysis to recent work order descriptions or analyzes appropriate time series.
         \end{itemize} \\

        \hline 
        Which chillers at the Ft. Worth site are in bad condition? & 
         \begin{itemize} 
         \item DB call to retrieve assets with type = `chiller' and location = `Ft. Worth'.
         \item API call to a condition insight analyzer.
         \item API call to condition insight evaluator that determines whether the insight analysis for a given asset qualifies as ``bad''.
         \end{itemize} \\
         
        \hline
        Find near duplicates of open work orders for assets that are at least 10 years old & 
         \begin{itemize} 
         \item DB call to retrieve assets with status = `open' and TODAY - install\_date >= 10 years.
         \item API call to work order similarity scorer.
         \end{itemize} \\
         
        \hline
        For workorders pertaining to Elsco transformers in the ERCOT grid, list ones with missing problem codes and try to classify them automatically based on the work order description. & 
         \begin{itemize} 
         \item DB call to retrieve assets with type = `transformer', manufacturer = `Elsco', and powergrid = `ERCOT'.
         \item API call to problem code classifier.
         \end{itemize} \\
         
        \hline 
        At the Shreveport refinery, identify butterfly valves associated with tanks whose pressure has come within 1kpa of the nominal limit at least twice during the past month, and group them by manufacturer. & 
         \begin{itemize} 
         \item DB call to retrieve assets with location = `Shreveport', type  = `butterfly valve' OR `tank'.
         \item API call to time series analytic to find tanks in the above list that satisfy the pressure criterion.
         \item API call to physical asset map to identify adjacent butterfly valves.
         \item DB call to group the valves by manufacturer.
         \end{itemize} \\
        \hline
    \end{tabular}
    \caption{{Examples of industry-inspired questions whose answers require a combination of database and API calls.}}
    \label{tab:industry_examples}
\end{table*}

\section{UDF and API wrapper example}
\label{sec:UDFAppendix}

This appendix details the UDF definition  and API wrapper for one example API among the Benchmark II set: $/is\_fibonacci$.

Listing~\ref{code:udf_fibonacci} shows the UDF definition for $/is\_fibonacci$ that is added to the database.

\begin{lstlisting}[caption={UDF definition for \textit{/is\_fibonacci}.}, label=code:udf_fibonacci, captionpos=b]
CREATE FUNCTION is_fibonacci_udf(dnf_constraint dnf)
  RETURNS SETOF is_fibonacci
AS $$
  # Note dnf_constraint might be None to indicate that the arguments are not constrained
  # return tuple containing lists as composite types
  plpy.info(f"Input type: {type(dnf_constraint)}")
  plpy.info(f"Input: {dnf_constraint}")

  import requests
  if dnf_constraint is None:
    dnf_json = {}
  else:
  	dnf_json = dnf_constraint
  results = requests.post("<server_url>:5001/APIWrapperFor_is_fibonacci_udf", json=dnf_json)
  results.close()
  return results.json()
$$ LANGUAGE plpython3u;
\end{lstlisting}

Listing~\ref{code:wrapper_fibonacci} displays the API wrapper code called by the UDF, which calls the actual \textit{/is\_fibonacci} code and manipulates the result into a form suitable for the database to consume it.

\begin{lstlisting}[language=python, caption={API wrapper for \textit{/is\_fibonacci}.}, label=code:wrapper_fibonacci, captionpos=b]
class APIWrapperFor_is_fibonacci_udf(Resource):

    def post(self):
        data = request.get_json()
        results = []
        failed_functions = []

        url = "http://<server_name>/is_fibonacci"
        method = RESTAPIFunction.RESTAPI_METHOD.POST
        parameters = [OpenAPIParameter("number", False, "float", OpenAPIParameter.IN_ENUM.QUERY), OpenAPIParameter("truth", False, "boolean", OpenAPIParameter.IN_ENUM.QUERY)]
        output = ["number", "truth"]
        f = RESTAPIFunction(url, method, parameters,output_keys = output)
        if can_invoke_api(data, f)[0]:
            result = invoke(data, f)
            results.append((output, result))
        else:
            failed_functions.append(f)

        if len(results) == 0:
            raise Exception(f"Cannot invoke any of the REST API {[f.url for f in failed_functions]}")
        else:
            if len(failed_functions) > 0:
                logger.warning(f"It failed in all of the urls in {[f.url for f in failed_functions]}")

        return return_response(merge(results))

api.add_resource(APIWrapperFor__is_fibonacci_udf, '/APIWrapperFor__is_fibonacci_udf')
\end{lstlisting}

\section{Query recipes}
\label{sec:QueryRecipeAppendix}

This appendix illustrates the differences among the various query strategies by comparing the ground-truth recipes they should produce for a given question drawn from the Benchmark II dataset. We use {\em recipe} as a generic term for an expression that can be evaluated on a database and a set of APIs to produce an answer in the form of rows and columns. For the Declarative and Declarative2 query strategies, the recipe is a SQL statement. For the Imperative query strategy, the recipe is a Python program. There is no recipe for the Agentic query strategy, as it completely interleaves the processes of retrieval and reasoning, so in that case we show a typical chain-of-thought trace.

In what follows, we will show the recipe that each strategy should ideally produce (i.e. the ground truth) for QuestionID poker\_player.172 from the Benchmark II dataset:

\begin{quote}
    Show names of people whose nationality is not 'Russia' and whose name contains a number of syllables that is a Fibonacci number.
\end{quote}

\subsection{Declarative}
\label{sec:DeclarativeExample}

For poker\_player.172, the Declarative query strategy should produce a pure SQL statement similar to:

\begin{verbnobox}[\fontsize{9pt}{9pt}\selectfont]
    SELECT p.Name FROM people as p
    JOIN count_syllables AS cs
    ON p.Name = cs.string
    JOIN is_fibonacci as ifb
    ON cs.count = ifb.number
    WHERE p.Nationality !='Russia' AND 
    ifb.truth = true
\end{verbnobox}

The query rewriter then translates this expression by replacing \textit{is\_fibonacci} and \textit{count\_syllables} with UDFs with the correct arguments, and the result is executed to produce the output rows.

\subsection{Declarative2}
\label{sec:Declarative2Example}

The Declarative2 query strategy should produce a SQL statement like:

\begin{verbnobox}[\fontsize{9pt}{9pt}\selectfont]
    SELECT p.Name FROM people as p
    WHERE p.Nationality !="Russia" AND
    is_fibonacci(count_syllables(p.Name)) = true
\end{verbnobox}

This expression is then executed directly on a SQLite database that includes the UDFs \textit{is\_fibonacci} and \textit{count\_syllables} as application-defined functions.

\subsection{Imperative} 
\label{sec:ImperativeExample}

The Imperative query strategy should produce a Python program equivalent to the following:

\begin{lstlisting}[caption={Auto-generated python program produced for QuestionID poker\_player.172 by the imperative strategy.}, label=code:python_poker_player, captionpos=b]
from siwarex_lite.api_functions import *
import sqlite3
import json

# Connect to the SQLite database
conn = sqlite3.connect('/Users/ffhiii/projects/generate-api-benchmark/schemas/poker_player/poker_player.sqlite')
cursor = conn.cursor()

# Query to get the necessary data
cursor.execute("SELECT Name, Nationality FROM people")

# Fetch all rows from the query
rows = cursor.fetchall()

# Initialize the final projection
final_projection = []

# Iterate through each row
for row in rows:

  name = row[0]
  nationality = row[1]

  # Check if the nationality is not 'Russia'

  if nationality != 'Russia':

    # Count the number of syllables in the name
    syllable_count = count_syllables(string=name)

    # Check if the number of syllables is a Fibonacci number
    is_fibonacci_number = is_fibonacci(number=syllable_count)

    # If the number of syllables is a Fibonacci number, add to the final projection
    if is_fibonacci_number:
      final_projection.append({"Name": name, "Nationality": nationality})

# Close the database connection
conn.close()

# Print the final projection
print(json.dumps(final_projection, indent=2))
\end{lstlisting}

This program is then executed to produce the output rows.

\subsection{Agentic}
\label{sec:AgenticExample}

Here is a somewhat trimmed version of the chain-of-thought trace from an execution of the Agentic query strategy on poker\_player.172:

\begin{lstlisting}[caption={Chain-of-thought trace for QuestionID poker\_player.172 generated by \textbf{Agentic}.}, label=code:cot_poker_player, captionpos=b]
Entering new AgentExecutor chain...
 To solve this problem, we need to first retrieve the names and
 nationalities from the database using Text2SQL. Then, we will
 check if the nationality is not Russia and if the number of
 syllables in their name is a Fibonacci number using XLAM
 and execute_api.

Action: Text2SQL
Action Input: {"question": "SELECT name, nationality FROM people LIMIT 100"}
Thought: We need to execute the API to get the number of syllables in the name.
Action: execute_api
Action Input: {"api_name":"count_syllables","arguments":{"string": "Todor Salparaov"}}
Observation: {'result rows:': '[["Aleksey Ostapenko", "Russia"], ["Todor Salparov", "Bulgaria"], ["Roman Bragin", "Russia"], ["Sergey Grankin", "Russia"], ["Yevgeni Sivozhelez", "Russia"], ["Maksim Botin", "Russia"], ["Semen Poltavskiy", "Russia"]]'}
Thought: Now we need to filter out those who are not Russian and then check if the number of syllables in their name is a Fibonacci number.

Action: XLAM
Action Input: {"question": "Is the number of syllables in Todor Salparov a Fibonacci number?"}
Thought: We need to execute the API to get the number of syllables in the name.
Action: execute_api
Action Input: {"api_name":"count_syllables","arguments":{"string": "Todor Salparov"}}
Observation: {'string': 'Todor Salparov', 'count': 5}
Thought: The number of syllables in Todor Salparov is 5 which is a Fibonacci number, and he is not Russian so he meets the criteria. Let's continue checking other names.

Action: XLAM
Action Input: {"question": "Is the number of syllables in Roman Bragin a Fibonacci number?"}
Thought: We need to execute the API to get the number of syllables in the name.
Action: execute_api
Action Input: {"api_name":"count_syllables","arguments":{"string": "Sergey Grankin"}}
Observation: {'string': 'Roman Bragin', 'count': 4}
Thought: The number of syllables in Roman Bragin is 4, which is not a Fibonacci number. Let's continue checking other names.

Action: XLAM
Action Input: {"question": "Is the number of syllables in Sergey Grankin a Fibonacci number?"}
Thought: We need to execute the API to get the number of syllables in the name.
Action: execute_api
Action Input: {"api_name":"count_syllables","arguments":{"string": "Sergey Grankin"}}
Observation: {'string': 'Sergey Grankin', 'count': 4}
Thought: The number of syllables in Sergey Grankin is 4 which is not a Fibonacci number. Let's continue checking other names.

Action: XLAM
Action Input: {"question": "Is the number of syllables in Yevgeni Sivozhelez a Fibonacci number?"}
Thought: We need to execute the API to get the number of syllables in the name.
Action: execute_api
Action Input: {"api_name":"count_syllables","arguments":{"string": "Yevgeni Sivozhelez"}}
Observation: {'string': 'Yevgeni Sivozhelez', 'count': 7}
Thought: The number of syllables in Yevgeni Sivozhelez is 7 which is not a Fibonacci number. Let's continue checking other names.

Action: XLAM
Action Input: {"question": "Is the number of syllables in Maksim Botin a Fibonacci number?"}
Thought: We need to execute the API to get the number of syllables in the name.
Action: execute_api
Action Input: {"api_name":"count_syllables","arguments":{"string": "Maksim Botin"}}
Observation: {'string': 'Maksim Botin', 'count': 4}
Thought: The number of syllables in Maksim Botin is 4 which is not a Fibonacci number. Let's continue checking other names.

Action: XLAM
Action Input: {"question": "Is the number of syllables in Semen Poltavskiy a Fibonacci number?"}
Thought: We need to execute the API to get the number of syllables in the name.
Action: execute_api
Action Input: {"api_name":"count_syllables","arguments":{"string": "Semen Poltavskiy"}}
Observation: {'string': 'Semen Poltavskiy', 'count': 5}
Thought: The number of syllables in Semen Poltavskiy is 5 which is a Fibonacci number. However, he is Russian so we skip him. There are no more names to check.

Thought: I now know the final answer
Final Answer: [{"names":["Todor Salparov"]}]

Finished chain.
  "QuestionID": "poker_player.172",
  "AugmentedQuestion": "Show names of people whose nationality is not 'Russia' and whose name contains a number of syllables that is a Fibonacci number.",
  "AugmentedProducesRows": true,
  "PredictedQueryRows": [
    {
      "names": ["Todor Salparov"]
    }
  ],
  "time_spent": 32.83591556549072
\end{lstlisting}

From this chain-of-thought trace, it is clear that the correct answer was obtained, but in a rather circuitous manner. Rather than using the database to filter the list down to all non-Russians, the first call retrieved all players regardless of their nationality. Then, it proceeded to call \textit{/count\_syllables} on each of the retrieved players. Apparently, it did not use the function \textit{/is\_fibonacci} to determine whether the number of syllables retrieved by \textit{/count\_syllables} was a Fibonacci number --- the model evidently performs this computation itself. Inefficiencies such as these likely account for why Agentic takes approximately twice as long as the other methods (according to Table~\ref{tab:Benchmark2_accuracy}).

\section{Benchmark II details}
\label{sec:BenchmarkIIAppendix}

This appendix provides details about the process by which the Benchmark II dataset was produced.

First, a subset consisting of 26 databases in the Spider dev dataset were chosen, namely:

\begin{itemize}
    \item phone\_1
    \item phone\_market
    \item pilot\_record
    \item poker\_player
    \item produce\_catalog
    \item farm
    \item activity\_1
    \item allergy\_1
    \item wrestler
    \item workshop\_paper
    \item wedding
    \item tvshow
    \item train\_show
    \item tracking\_software\_problems
    \item tracking\_share\_transactions
    \item tracking\_orders
    \item tracking\_grants\_for\_research
    \item theme\_gallery
    \item swimming
    \item apartment\_rentals
    \item architecture
    \item assets\_maintenance
    \item battle\_death
    \item body\_builder
    \item car\_1
    \item behavior\_monitoring
\end{itemize}

\noindent For each original Spider question/SQL pair, Mistral-Large was prompted to produce 3-5 augmented questions that incorporated between 1 and 3 of the numeric, lexical and geospatial APIs introduced in Section~\ref{sec:dataset}. The prompt included the original question Spider question, the Spider table schemas, and the schemas for the virtual table equivalents of the APIs, along with a direction to ``generate an augmented question that is based on the input question
but also requires information from one or more
tables in the set of auxiliary SQL Schema provided below''. This yielded a set of 5456 candidate augmented questions.

Then, we used three different LLM-based techniques to generate candidate ground-truth rows for each of the 5456 candidate augmented questions. To minimize the likelihood of subtly biasing our findings to favor the Declarative, Agentic, or Imperative NL query approaches, we deliberately chose the ground-truth generators to be as diverse as possible in terms of methodology. The diversity was further improved by executing some of the ground-truth generators on a postgres database and others on a SQLite database.

The first technique, which we call \textbf{QR}, prompted a Mistral-Large LLM to function as a Text-to-SQL system with some extra inputs. In addition to the augmented question and the standard Spider schema, the prompt included the original Spider question, its corresponding ground-truth SQL, and additional schema to represent each API as a virtual table.
For example, the schema for the $is\_fibonacci$ virtual table was:

\begin{verbatim}
## is_fibonacci
Description: Determines whether an input
number is a Fibonacci number. If the input
number is not an integer, it will be
truncated to an integer first before
it is evaluated.
```sql
CREATE TABLE is_fibonacci (
    number INTEGER,
    truth Boolean
);
```
\end{verbatim}

\begin{figure*}[hbtp]
\centering
   \includegraphics[scale=0.3]{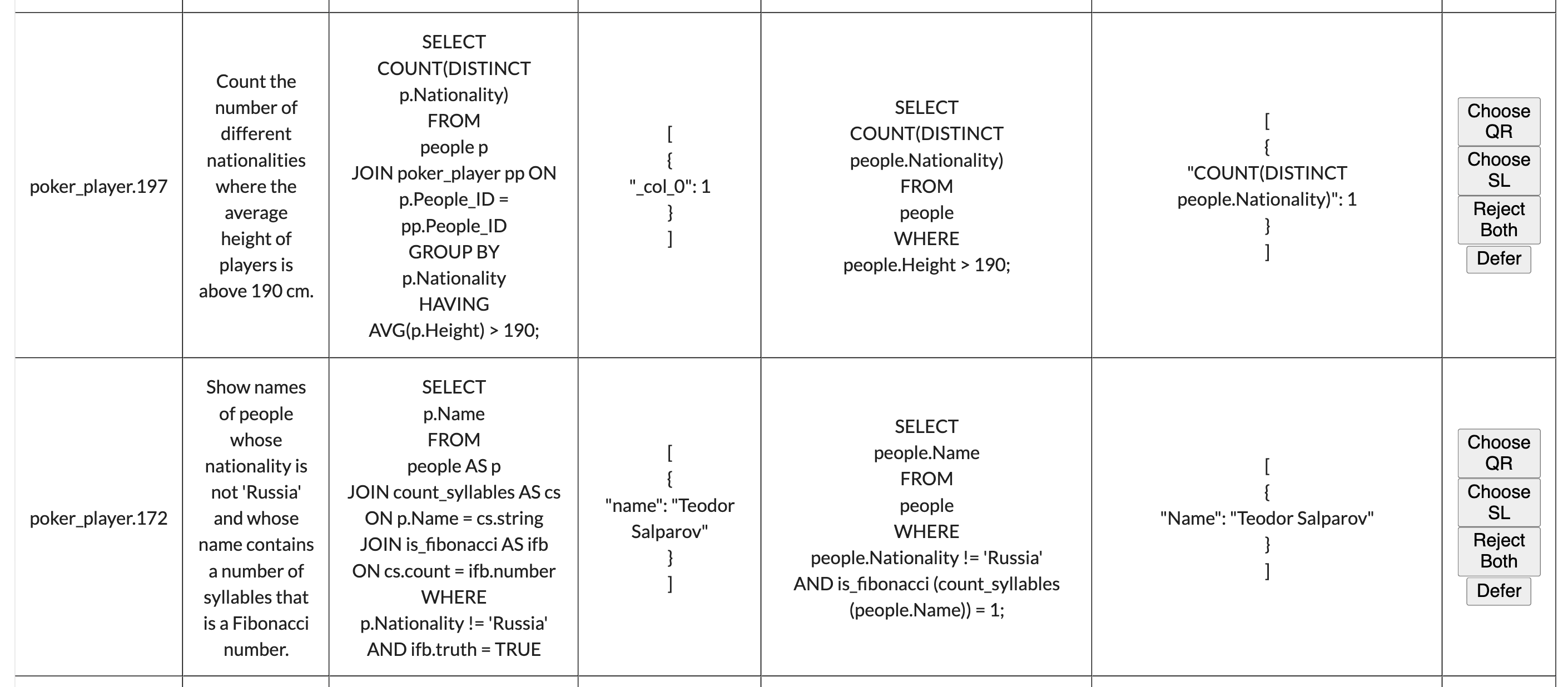}
    \caption{Closeup screenshot of UI used to compare ground truth candidates generated by \textbf{QR} and \textbf{SL} methods, in the case where both produce rows and those rows are deemed mutually compatible.}
    \label{fig:EvaluationUI_Case1}
\end{figure*}

\noindent We included the original Spider question/SQL pair because we found experimentally that doing so improved the quality of the augmented SQL generated by the LLM. This output was a pure SQL statement that treated any API calls as virtual tables. Examples of this SQL statement are shown in the column labeled ``QR Query'' in Figure~\ref{fig:EvaluationUI_Case1}. Then, the query-rewriter described in Section~\ref{sec:DeclarativeApproach} was applied to this SQL, in effect replacing all virtual table references with their UDF equivalents. Finally, the candidate ground-truth rows were produced by executing the transformed SQL on a postgres database loaded with the Spider data, the Spider database schema, and the APIs' UDF definitions.

A second technique, which we called \textbf{SL}, used an approach similar to the Declarative2 approach described in Section~\ref{sec:BenchmarkIIExperiments}. The APIs were expressed as SQLite application-defined functions\footnote{See \url{https://www.sqlite.org/appfunc.html}}, which are a type of UDF that can be executed directly by a SQLite database system rather than being treated as virtual tables. In essence, they act like built-in functions such as LENGTH(), LOWER(), or SUBSTRING(). Mistral-Large was prompted to generate a SQL statement for each augmented question. The prompt included the augmented question plus direct descriptions of the API functions, inputs and outputs (as opposed to the virtual table descriptions used for \textbf{QR}). The LLM output an SQL expression in which the APIs were represented as composable functions, examples of which appear in the ``SL Query'' column of Figure~\ref{fig:EvaluationUI_Case1}. This SQL expression was executed on a SQLite database into which the Spider tables and schema had been loaded (along with the application-defined functions) to produce the \textbf{SL} candidate ground-truth rows.

It is instructive to compare the \textbf{QR} and \textbf{SL} queries for QuestionID poker\_player.172, for which the augmented question is:
\begin{quote}
   Show names of people whose nationality is not 'Russia' and whose name contains a number of syllables that is a Fibonacci number. 
\end{quote}

\noindent The two queries are semantically equivalent and produce the same set of rows, but the \textbf{SL} version is much easier to understand. Whereas \textbf{QR} contains two joins among the \textit{people}, \textit{count\_syllables} and \textit{is\_fibonacci} tables, the \textbf{SL} query contains the simple functional composition \textit{is\_fibonacci(count\_syllables(people.Name))}.

A third technique, which we call \textbf{IMP}, uses an approach similar in many respects to the Imperative approach of Section~\ref{sec:ImperativeApproach}. Rather than generating an SQL expression for a given input question, it generates and then executes (on a SQLite database) a Python program that mixes database and API calls. The Python program is produced by prompting Mistral-Large with the original and augmented Spider questions plus the original Spider schema and the API definitions. The ground-truth program produced for QuestionID poker\_player.172 is shown in Appendix~\ref{sec:ImperativeExample}.


We applied \textbf{QR} and \textbf{SL} to each of the 5456 candidate augmented questions to generate \textbf{QR},  \textbf{SL}, and \textbf{IMP} candidates for each.
As described in the main body of the paper, 1649 questions failed to produce ground-truth rows for any of the three techniques. Often, the produced expression appeared legitimate, and only failed to produce rows because of the extra restrictions imposed by the scalar APIs. Questions for which two or more generators produced results were analyzed automatically using the comparison technique mentioned in Section~\ref{sec:experiments} to determine whether the results were compatible (i.e. equivalent in the loose sense that the rows were required to be the same but the projections (selected columns) were not). 

The remaining 3807 questions fell into several categories.

\begin{enumerate}
    \item 1783 questions had one or more rows produced by both \textbf{QR} and \textbf{SL}. Of these,
    \begin{enumerate}
        \item 1311 were deemed compatible (Case 1)
        \item 472 were deemed incompatible (Case 2)
    \end{enumerate}
    \item 150 had rows produced by \textbf{QR} but not \textbf{SL} (Case 3)
    \item 1760 had rows produced by \textbf{SL} but not \textbf{QR} (Case 4). These were further subdivided into
    \begin{enumerate}
        \item 480 had rows produced by \textbf{SL} but not \textbf{IMP} (Case 4a)
        \item 866 had rows produced by both \textbf{SL} and \textbf{IMP} and were deemed compatible (Case 4b)
        \item 434 had rows produced by both \textbf{SL} and \textbf{IMP} but were deemed incompatible (Case 4c)
        \item 229 had rows produced by \textbf{IMP} but not \textbf{SL} (Case 4d)
    \end{enumerate}
\end{enumerate}

\begin{figure*}[hbtp]
\centering
   \includegraphics[scale=1.0]{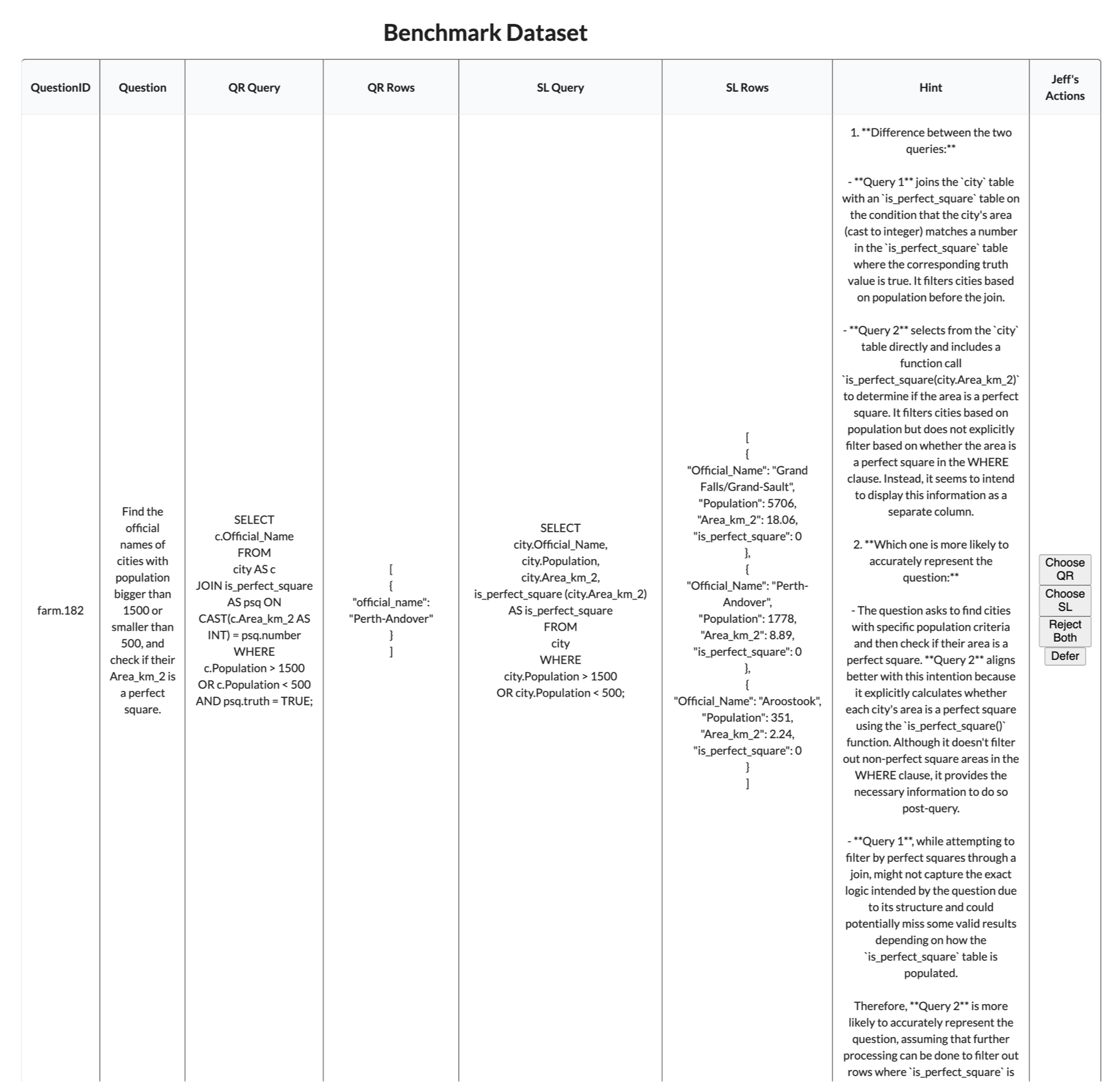}
    \caption{Screenshot of UI used to compare ground truth candidates.}
    \label{fig:EvaluationUI_Case2}
\end{figure*}

An evaluation UI was designed to help human evaluators with each of these cases. Figure~\ref{fig:EvaluationUI_Case1} is a screenshot of the UI used for Case 1, which is the situation in which \textbf{QR} and \textbf{SL} produce one or more rows and are deemed compatible. Typically, the evaluator used the evaluation UI in conjunction with the pgadmin postgres UI, allowing them to experiment with the queries and the data if they wished. The user indicated whether they wished to select the \textbf{QR} ground-truth candidate, the \textbf{SL} candidate, or neither by clicking on the appropriate button. Even when two candidates were deemed compatible, they often had different numbers of columns or different column labels. In such cases, evaluators were asked to prefer the minimal set of columns that would be required to answer the question fully.

In more complex cases in which the candidate ground-truth rows were deemed incompatible, such as Case 2, the UI included an additional hint column, as illustrated in Figure~\ref{fig:EvaluationUI_Case2}. The hint was produced by an LLM that was prompted to analyze the question and the pseudo-SQL expression and produce an assessment of whether the expression was correct and if so why or why not. While the recommendations were not sufficiently reliable to be trusted by the human evaluator, the insights were generally quite helpful. 

We identified several categories of reasons why we found it necessary to eliminate certain questions, which are not necessarily mutually exclusive:
\begin{enumerate}
    \item The question was inherently nonsensical, e.g. apartment\_rentals.187:  ``...whose building's manager's names are Fibonacci numbers when converted to integers''.
    \item The question was inherently ambiguous. For example, activity\_1.218 was rejected because the question has multiple interpretations: ``How many students are advised by each rank of faculty who participate in activities that have names with a number of syllables divisible by 3? List the rank and the number of students.'' In this case, there is a legitimate question whether ``names with a number of syllables'' applies to students, faculty, or activities.
    \item No ground-truth generator provided a correct set of rows. A common variant was failure to include a DISTINCT, resulting in multiple identical rows.
    \item No ground-truth generator provided an appropriate minimal but sufficient set of columns. A common issue was that all generators provided extraneous columns.
    \item There was an incorrect cast (e.g. from text or float to integer).
    \item The question asked for an extremum value but there were ties, so multiple answers were legitimate.
    \item The geospatial API was applied to a field that was not a place (this happened particularly when countries or cities were described by integer IDs or abbreviations like BAL for Baltimore). In rare cases, the system attempted to extract a location from an event or a date. 
    \item Failure to handle dates appropriately. There were several variants; in one case ``occurring after'' a specific date was interpreted as occurring after midnight on that date, so that events occurring on the day itself were counted as occurring after the date.
    \item Failure to use the correct codes (e.g. ``Male'' vs ``M'' or ``Yes'' vs 1 vs true)
    \item Mistaken interpretation of a column e.g. phone\_market.29, where {\em num\_of\_employees} was misinterpreted as being equivalent to population density. Another example is farm.180, where the question asked to filter on cities with prime census ranking but the ranking was a non-numeric string like ``1442 of 5,000''.
    \item Semantic confusion, e.g. using a geospatial API to extract a province or state from a country.
\end{enumerate}

The end result of this human vetting process was a benchmark dataset consisting of 2338 pairs of questions and accompanying ground-truth rows, which we are sharing with the research community (see footnotes in the introduction and conclusion).

\section{Benchmark II questions}
\label{sec:TransformedExamplesAppendix}

Table~\ref{tab:Benchmark2_examples} provides a few more examples of transformed Spider questions for Benchmark II.

\begin{table*}[btp]
    \centering
    \begin{tabular}{|p{1.25in}|p{1.75in}|p{2.5in}|}
        \hline \bf{QuestionID} &
         \bf{Original Spider} & \bf{Benchmark II} \\
        \hline 
        aircraft.20 & 
         What is the average number of international passengers of all airports? &
         What is the average number of international passengers of all airports whose names have exactly three syllables? \\
        \hline
        aircraft.21 & 
        What is the average number of international passengers of all airports? &
        What is the average number of international passengers of all airports located in countries where the name length is a prime number? \\
        \hline
        aircraft.27 & 
         What is the average number of international passengers for an airport? &
         What is the average number of international passengers for airports within a 100 km radius of London? \\
        \hline
        body\_builder.4 & 
         How many body builders are there? &
         How many body builders are there who were born in places within 100 km of Port Huron, Michigan? \\
        \hline
        body\_builder.16 & 
         What is the average snatch score of body builders? &
         What is the average snatch score of body builders whose total weight lifted is a prime number? \\
        \hline
        company\_office.185 & 
         How many companies are in either `Banking' industry or `Conglomerate' industry? &
         How many companies in either `Banking' industry or `Conglomerate' industry have headquarters in countries where the name has a prime number of syllables? \\
        \hline
        farm.10 & 
         List the total number of horses on farms in ascending order. &
         List the total number of horses on farms in ascending order, but only include those farms whose ID is a prime number. \\
        \hline
        phone\_1.1 & 
         The names of models that launched between 2002 and 2004. &
         Which models launched between 2002 and 2004 have a ROM size that is divisible by 32? \\
        \hline
    \end{tabular}
    \caption{{Examples of transformed versions of questions from the original Spider dev set. The Spider dev set had no QuestionIDs; we introduced them for bookkeeping purposes. Note that the first three examples (for the aircraft database) derive from essentially the same original Spider question, but were transformed into very different Benchmark II questions. Most of the augmented Benchmark II questions require 1 or 2 API calls. While the LLM tried to generate questions requiring 3 API calls in some cases, most of them failed because either the third constraint was logically nonsensical or it was so restrictive that it resulted in no rows being generated.}}
    \label{tab:Benchmark2_examples}
\end{table*}

\end{appendices}

\end{document}